\documentclass[aps,prd,twocolumn,superscriptaddress]{revtex4-2}
\usepackage{graphicx}
\usepackage{float}
\usepackage{amsmath}
\usepackage{amsfonts}
\usepackage{color}
\usepackage{bm}
\usepackage{bbm}
\usepackage{array}
\usepackage{amssymb}
\usepackage{microtype}
\usepackage[english]{babel}
\usepackage{soul}
\usepackage{siunitx}
\usepackage{placeins}
\usepackage[colorlinks=true,
    linkcolor=red,
    urlcolor=magenta,
    citecolor=blue,
    anchorcolor=blue]{hyperref}
\usepackage[nameinlink]{cleveref}
\crefname{figure}{Fig.}{Figs.}
\usepackage[caption=false]{subfig}

\begin{document}
	\title{Enhancement of Proton Acceleration via Geometric Confinement in Near Critical Density-filled Targets }
    \author{Cheng-Qi Zhang}
	\affiliation{
	Key Laboratory of Beam Technology of the Ministry of Education, and School of Physics and Astronomy, Beijing Normal University, Beijing 100875, China}

	\author{Yang He }
	\affiliation{Xinjiang Key Laboratory of Solid State Physics and Devices,
	School of Physics Science and Technology, Xinjiang University, Urumqi 830017, China}
	\author{Mamat Ali Bake }
	\affiliation{Xinjiang Key Laboratory of Solid State Physics and Devices,
	School of Physics Science and Technology, Xinjiang University, Urumqi 830017, China}
		\author{Bai-Song Xie}\thanks{Corresponding author. Email: bsxie@bnu.edu.cn}
	\affiliation{
	Key Laboratory of Beam Technology of the Ministry of Education, and School of Physics and Astronomy, Beijing Normal University, Beijing 100875, China}
	\begin{abstract}
	High-quality proton beams generated by laser-plasma interactions are of significant interest for applications ranging from tumor therapy to fast ignition in inertial confinement fusion. However, simultaneously achieving high energy coupling efficiency and beam collimation remains a challenge. In this work, we investigate the enhancement of proton acceleration via geometric confinement in Near-Critical Density (NCD) plasma-filled micro-structured targets using two-dimensional particle-in-cell (PIC) simulations. To optimize laser-to-particle energy transfer, we systematically compared various target configurations, such as rectangular tubes, hybrid funnels, and straight cones. Our results reveals that increasing geometric complexity does not necessarily translate to superior acceleration performance. Instead, the relatively simple NCD-filled straight-cone target outperforms more complex hybrid geometries, achieving a maximum proton cutoff energy of 181.7 MeV and a reduced divergence of approximately $12^{\circ}$ at a laser intensity of $5.5 \times 10^{20}$ W/cm$^2$. This enhancement is attributed to the synergistic effect of relativistic laser self-focusing within the NCD channel and the strong spatial confinement of hot electrons by the conical walls. Furthermore, we identify a unique double-peak structure in the temporal evolution of the electron energy, which serves as a signature of sustained electron refluxing. This refluxing mechanism maintains a robust sheath field over an extended duration, driving the superior acceleration. The proposed target design offers a robust pathway for generating high-flux, high-energy proton beams suitable for next-generation high-repetition-rate laser facilities.
	\end{abstract}
	\maketitle
	
\section{Introduction}
High-energy proton beams have attracted increasing interest due to their potential in diverse fields, ranging from proton radiography \cite{borghesi2002macroscopic} to medical ion therapy \cite{linz2007will} and fast ignition in inertial confinement fusion \cite{roth2001fast}. Unlike conventional accelerators constrained by massive scale and cost, laser-driven ion sources offer a compact and cost-effective alternative \cite{macchi2013ion}. These sources are regarded as a primary driver for the next generation of high-repetition-rate laser facilities, providing the ultra-fast and energetic beams required for extreme physical processes \cite{prencipe2017targets}.

For applications such as tumor therapy, the Bragg peak effect allows charged particles to release most of their kinetic energy at the end of their trajectory\cite{wilson1946radiological}, requiring therapeutic proton beams to reach 70--250 MeV for sufficient deep tissue penetration\cite{mohan2017proton}. Beyond energy, beam qualities such as divergence is another critical factor for clinical viability\cite{toncian2006ultrafast}.

To address this, strategies ranging from optical wavefront shaping using Laguerre-Gaussian (LG) beams \cite{allen1992orbital, brabetz2015laser, pang2025enhanced} to plasma collimation utilizing water vapor targets \cite{streeter2025stable} have been explored. While these recent advances can reduce divergence to the degree level under specific conditions\cite{wang2025enhanced, streeter2025stable}, they often entail trade-offs such as laser energy loss or increased experimental complexity. Consequently, developing a robust target scheme that simultaneously ensures high coupling efficiency and beam collimation remains a key objective.

\begin{figure*}[t]
 \centering
\includegraphics[keepaspectratio, width=1.08\textwidth]{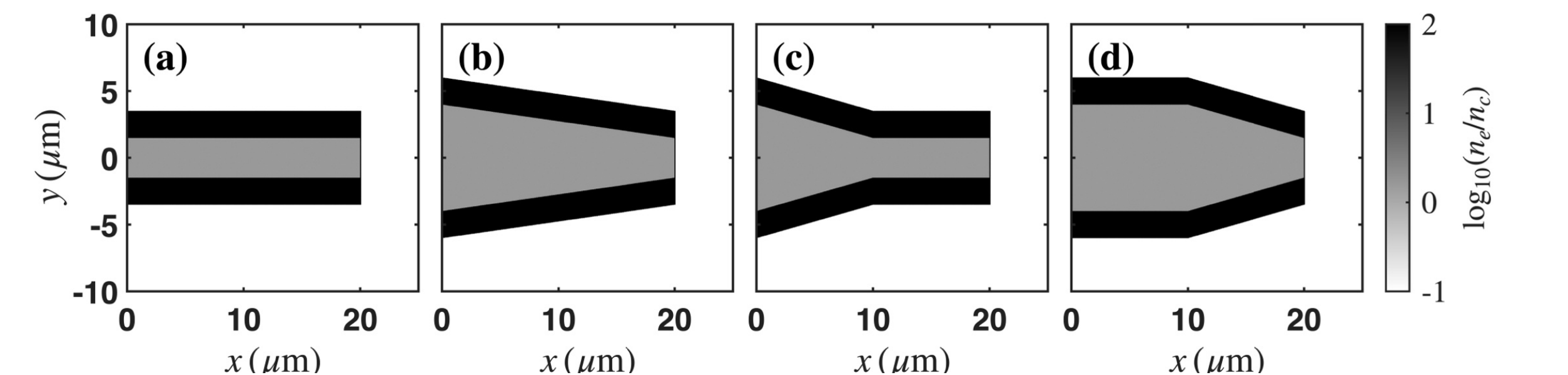}  
  \caption{Initial electron density distribution for representative PIC simulation cases: (a) rectangular-grooved target, (b) straight-cone target, (c) hybrid funnel target with a cone-front geometry, and (d) hybrid funnel target with a cone-back geometry. The laser pulse is incident from the left boundary along the $x$-axis. The primary 60 nm hydrogen target is positioned at $x = 20\,\mu\text{m}$. }  
  \label{fig:Fig.1}  
\end{figure*}

The physics of laser-proton acceleration varies with laser and target parameters. Target Normal Sheath Acceleration (TNSA)\cite{wilks2001energetic}is recognized for its experimental robustness \cite{passoni2010target} but is limited by large divergence ($\sim 20^\circ$) and broad, exponentially decreasing spectra\cite{borghesi2004multi, brack2020spectral}. Conversely, Radiation Pressure Acceleration (RPA) \cite{esirkepov2004highly} and Relativistic Induced Transparency (RIT) \cite{yin2006gev} offer higher efficiency and energy potential, yet require ultra-high laser contrast or nanometer-scale targets that are challenging to implement stably \cite{robinson2008radiation, bulanov2016radiation, kim2016radiation}. Other mechanisms such as Shock Wave Acceleration (SWA)\cite{silva2004proton}, Magnetic Vortex Acceleration (MVA)\cite{bulanov2010generation,li2022enhancement} and hybrid strategies have also been explored\cite{kim2016radiation,higginson2018near,shou2025laser,shou2025proton,gupta2025enhanced}. Notably, Ziegler et al. proposed a "cascaded" regime that synergizes RPA, RIT, and TNSA to reach a record maximum proton energy of 150 MeV to date\cite{ziegler2024laser}. Nevertheless, reaching higher energy thresholds in a controllable manner remains challenging due to limited interaction time and incomplete energy coupling \cite{vallieres2021enhanced}.

 To overcome this, artificial intelligence (AI) and machine learning (ML) have been integrated to autonomously discover optimal configurations for the multi-dimensional parameter space\cite{loughran2023automated,mcqueen2025neural,desai2025applying}. Simultaneously, target engineering has emerged as another crucial approach to optimize particle distributions. Structured targets, such as nanometer foils \cite{shou2025proton}, helical coils \cite{liu2024synergistic}, micro-cones \cite{ebert2021targets,khan2023enhanced}, multilayers \cite{tong2018target}, and nanowires \cite{vallieres2021enhanced}, have been employed to enhance laser absorption through geometric effects.

Beyond structural modifications, the integration of Near-Critical Density (NCD) plasma as an auxiliary medium has shown exceptional promise. NCD media can provide relativistic pulse shaping and steepen the laser front \cite{bin2015ion}. For example, Sharma demonstrated that NCD hydrogen gas can serve as a debris-free source for high-repetition-rate acceleration \cite{sharma2018high}, while Prencipe et al. utilized cluster-assembled carbon foams to achieve significant gains in proton cut-off energies \cite{prencipe2021efficient}.More recently, Horný and Doria demonstrated that an NCD front layer can function as a plasma lens, enhancing laser intensity via relativistic self-focusing and increasing the population of high-energy protons\cite{horny2025multi}.

Traditionally, the precise fabrication of complex micro-structures or the exact morphological control of NCD foams posed significant experimental challenges, particularly to meet the demanding requirements of high-repetition-rate operations \cite{prencipe2017targets}. However, the recent advent of advanced 3D-printing technologies, such as two-photon polymerization (2PP), has revolutionized target engineering. These sub-micrometer additive manufacturing techniques enable the fine control of target morphology, strut thickness, and average density at the micro-scale \cite{christ2025two}. Consequently, the experimental realization and high-power laser irradiation of intricate, reproducible foam-based micro-targets have recently been demonstrated to be highly feasible \cite{cipriani2026experimental}. 

Recent studies have explored the integration of NCD plasmas within micro-structures to optimize specific mechanisms, such as magnetic vortex acceleration (MVA) in open-ended conical targets \cite{li2022enhancement}. However, the physical dynamics change  when these structures are coupled with a solid foil. The synergy between NCD-driven volumetric heating and geometric spatial confinement in foil targets remains unexplored. Specifically, it is not yet understood how such confinement modulates the evolution of charge-separation fields and the effective acceleration duration, nor is it clear how these coupled dynamics can concurrently enhance proton energy and suppress beam divergence.

In this work, we present a systematic numerical investigation into the proton acceleration dynamics within NCD-filled micro-structured targets. Using particle-in-cell (PIC) simulations\cite{tskhakaya2007particle}, we explore configurations ranging from simple rectangular apertures to hybrid funnel structures. We demonstrate that a straight-cone target filled with NCD plasma can significantly boost the proton cut-off energy to 181.7 MeV, with a reduced divergence of $12^\circ$. Our analysis reveals a unique double-peak electron energy evolution, where a secondary peak at approximately 400 fs signifies a critical energy replenishment process driven by electron refluxing within the NCD reservoir. This research provides a robust framework for designing high-flux, high-energy proton sources tailored for the next generation of high-repetition-rate laser facilities.

\section{simulation setup}\label{SecII}
The proton acceleration process is investigated using the relativistic PIC code EPOCH \cite{arber2015contemporary}. Given the expansive parameter space involved, we prioritize two-dimensional simulations to achieve a comprehensive and computationally efficient scan of the geometric and plasma configurations. This strategy allows for robust target optimization while retaining the essential physics. Previous studies have demonstrated that 2D simulations reliably reproduce the scaling trends of ion cut-off energies\cite{babaei2017rise}. Therefore, this work serves as a foundational feasibility study based on 2D modeling, paving the way for targeted three-dimensional investigations in the future.

\begin{figure*}[t]
 \centering
 \includegraphics[keepaspectratio, width=1.035\textwidth]{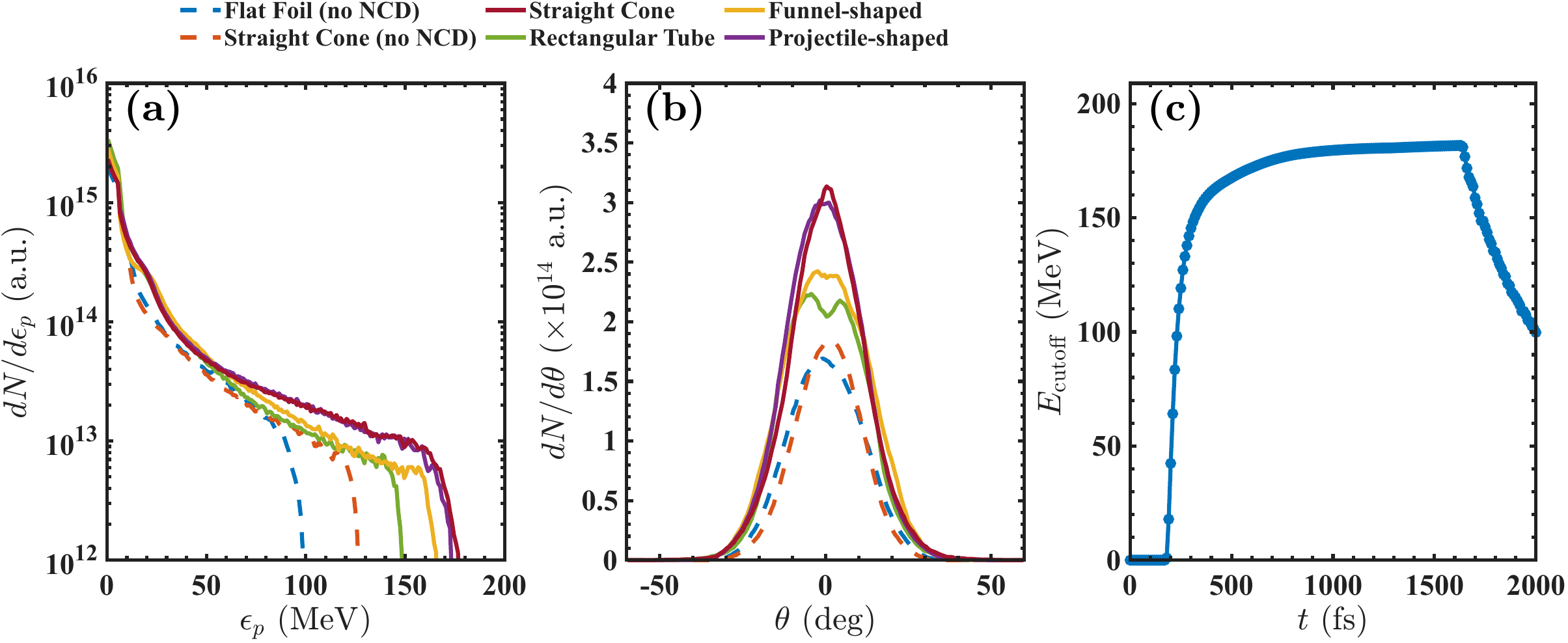}
 \caption{
    \protect\label{fig:2a}(a) Proton energy spectra at $t = 680\,\text{fs}$, 
    \protect\label{fig:2b}(b)Angular distribution ($dN/d\theta$) of protons with energy $\epsilon_p > 10\,\text{MeV}$ at t = 680fs,
     \protect\label{fig:2c}(c) Temporal evolution of proton cutoff energy in the straight cone target. }
 \label{fig:2}
\end{figure*}

The laser parameters are specifically chosen to represent the current operational capabilities of major Petawatt-class facilities, such as the Shanghai Superintense Ultrafast Laser Facility (SULF)\cite{gan2021shanghai}, the Draco-PW laser\cite{schramm2017first}, and the Astra Gemini facility at the Rutherford Appleton Laboratory\cite{scullion2017polarization}. As shown in \cref{fig:Fig.1}, a $p$-polarized Gaussian pulse with its electric field directed along the $y$-axis, a central wavelength of $\lambda = 0.8\,\mu\text{m}$, and a temporal full-width at half-maximum of $40\,\text{fs}$ is incident from the left boundary and propagates along the $x$-axis. The laser is focused to a spot size of $r_0 = 3.0\,\mu\text{m}$, yielding a peak intensity of $5.5 \times 10^{20}\,\text{W/cm}^2$.
To systematically optimize the acceleration efficiency, the target configurations are categorized into two evolutionary groups. The study begins with fundamental geometries, including a baseline flat foil and basic micro-structures such as rectangular tube, parabolic target and straight cone. The influence of the aperture width for the rectangular configurations is also investigated. Subsequently, composite "hybrid" targets are introduced, featuring funnel-shaped (cone-to-rectangle) and projectile-shaped (rectangle-to-cone) designs. This hierarchical approach aims to determine whether the geometric complexity of hybrid structures offers a significant advantage over the optimized fundamental rectangular targets.

The simulation domain is discretized into $5500 \times 1000$ cells and the target consists of a 60 nm thick hydrogen foil with an initial electron density of $100n_c$, where $n_c$ is the critical density. The foil thickness was set to $d = 60~nm$ based on a comprehensive parameter scan. This choice falls within the optimal transparency-enhanced regime identified in recent studies. For instance, Khan and Saxena demonstrated that for micro-structured targets, reducing the rear wall thickness to the nanometer scale triggers RIT near the laser peak, leading to a drastic boost in proton cutoff energy\cite{khan2025effect}. In our setup, $60~nm$ represents the ideal balance between sustaining the sheath field and leveraging transparency-driven acceleration. A thinner target tends to become transparent too early due to the additional volumetric heating from the NCD plasma, leading to a premature collapse of the accelerating sheath. Conversely, a thicker target suppresses the RIT enhancement. In hybrid cases, the micro-structure is filled with a uniform NCD plasma with $n_e = 1.0n_c$. 

The structural integrity of the ultra-thin target is highly sensitive to the laser prepulse. To maintain the prepulse intensity well below the typical ionization threshold of solid targets, an ultra-high laser contrast on the order of $10^{10}$ to $10^{11}$ is strictly required according to our peak intensity. Modern Petawatt-class laser facilities routinely achieve such contrast levels by employing single or double plasma mirror systems \cite{levy2007double,choi2020highly,czapla2025renewable}. Furthermore, in our proposed composite configuration, the NCD medium filling the micro-structure naturally acts as a supplementary physical buffer. As demonstrated in experimental studies on foam-based targets \cite{prencipe2021efficient,sgattoni2012laser,passoni2014energetic}, the low-density NCD material is preferentially ionized by the rising edge of any residual prepulse. This interaction effectively absorbs and scatters the prepulse energy, thereby mitigating the premature hydrodynamic expansion of the 60 nm primary foil and preserving its structural integrity for the subsequent main pulse interaction. The proposed composite target design falls entirely within the feasible operational regime of current state-of-the-art experimental setups.

To ensure numerical stability and minimize stochastic noise, 50 particles per cell are assigned to both the H-foil and the NCD regions. Open boundary conditions are applied to both particles and fields. 
\begin{figure*}[t]
 \centering
 \includegraphics[keepaspectratio, width=1.035\textwidth]{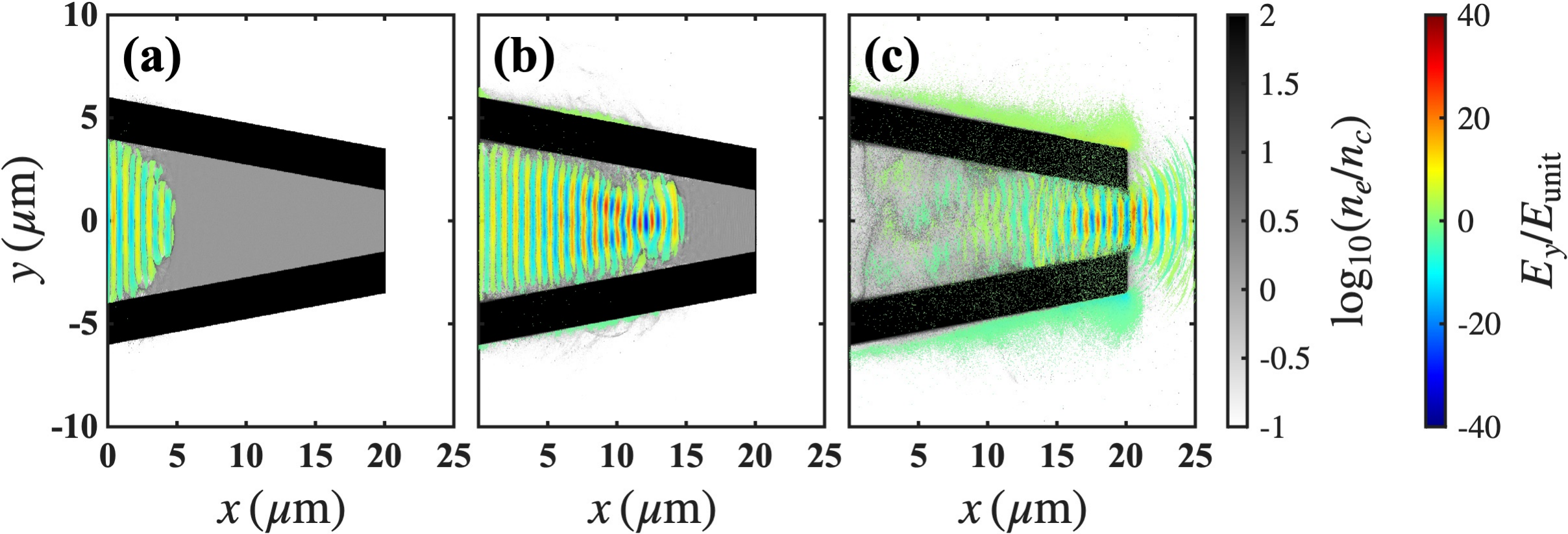}
 \caption{
    Spatiotemporal evolution of the laser transverse electric field and electron density distribution in the NCD-filled straight cone target at (a) $t=100\,\text{fs}$, (b) $t=150\,\text{fs}$, and (c) $t=200\,\text{fs}$. }
 \label{fig:3}
\end{figure*}

\section{result and discussion}\label{SecIII}

In this section, we present a systematic analysis of the proton acceleration dynamics across various target configurations, focusing on the interplay between NCD plasma and micro-structured geometric confinement. We evaluate the evolution of charge-separation fields and the resulting particle distributions to identify the optimal acceleration regime. The following discussion begins with a comparative performance benchmark of various target archetypes.

Prior to evaluating the conical and hybrid targets, we established an optimized baseline for the rectangular tube configuration. The choice of the aperture width ($D$) is governed by a trade-off between laser energy coupling and geometric focusing. As identified by Snyder et al.\cite{snyder2016enhancement}, an optimal inner diameter exists (typically around $4\lambda$) that maximizes the laser intensification factor by balancing diffraction effects against plasma clogging. Similarly, studies on micro-tube targets suggest that matching the aperture to the laser focal spot radius is crucial for maintaining beam quality\cite{khan2023enhanced}. To verify this in our specific NCD-filled setup, we performed a parameter scan comparing a wide aperture ($D = 8~\mu m$) and a narrow aperture ($D = 3~\mu m$). While the wide-aperture rectangular target achieved a high cutoff energy of 164 MeV due to maximum laser energy uptake, it suffered from severe beam divergence ($>18^{\circ}$ ), rendering it less suitable for applications requiring collimated beams. In contrast, the $3~\mu m$ aperture provided a highly collimated beam, albeit with lower energy due to laser clipping. Consequently, to rigorously test the capability of the conical geometry to significantly boost energy without compromising beam quality, we selected the $3~\mu m$ rectangular target as the primary high-quality benchmark for the comparisons presented in \Cref{fig:2}.

The performance of various target configurations is evaluated at $t = 680\,\text{fs}$, a temporal milestone where the proton acceleration process reaches a quasi-steady state. \cref{fig:2}(a) illustrates the proton energy distribution for the investigated geometries. We first assess the influence of geometric confinement by comparing the vacuum targets (dashed lines). The empty straight cone (orange dashed line) extends the maximum proton cutoff energy from approximately $100\,\text{MeV}$ to $130\,\text{MeV}$. This improvement is attributed to the spatial confinement provided by the conical walls, which facilitates electron focus and reinforces the longitudinal sheath field, consistent with the guiding mechanisms described by Honrubia et al.\cite{honrubia2017intense}.

The integration of NCD plasma increases this baseline performance. When comparing the NCD-filled straight cone (solid red line) with its empty-cone counterpart (orange dashed line), $\epsilon_{p}$ increases significantly to $176\,\text{MeV}$. This nearly 80\% enhancement over the flat foil baseline is accompanied by a marked increase in particle flux within the high-energy tail ($> 100\,\text{MeV}$). To isolate the optimal geometry, various other structured targets including rectangular tube, funnel-shaped, and projectile-shaped were benchmarked as well. While all NCD-assisted structures exhibit improved energy reach over the flat foil, the straight-cone configuration achieves the highest cutoff energy and particle counts.

\cref{fig:2b}(b) presents the angular distribution specifically for high-energy protons with $\epsilon_p > 10\,\text{MeV}$. The NCD-filled straight cone demonstrates superior beam collimation, evidenced by the highest peak spectral intensity and the narrowest divergence. The divergence of the straight cone is  constrained with a half-angle of approximately $12^{\circ}$, narrower than the profiles of the funnel-shaped or rectangular tube targets. While the projectile-shaped target shows a comparable peak flux, the straight cone maintains higher profile symmetry and a more concentrated axial emission.

The temporal evolution of the proton cutoff energy in the straight cone target is depicted in \cref{fig:2c}(c), which tracks the maximum proton energy over time. In the initial phase ($t < 500\,\text{fs}$), the cutoff energy surges rapidly with a steep gradient. In the subsequent phase ($500\,\text{fs} < t < 2000\,\text{fs}$), while the acceleration rate naturally moderates, the energy continues to rise exhibiting a sustained quasi-linear growth. This persistent acceleration ultimately propels the protons to a final cutoff energy of 181.7 MeV at the end of the simulation.

To elucidate the physical mechanism responsible for the enhanced acceleration, we investigate the spatiotemporal evolution of the laser pulse and plasma dynamics.  \Cref{fig:3} presents the transverse electric field $E_y$ and the electron density distribution for the NCD-filled Straight Cone target at three critical interaction stages. At the injection phase ($t=100\,\text{fs}$, \cref{fig:3}(a)), the laser pulse enters the NCD region before interacts with the cone walls, where the beam undergoes relativistic self-focusing. As the pulse propagates deeper ($t=150\,\text{fs}$, \cref{fig:3}(b)), the geometric confinement becomes dominant. The conical walls restrict the transverse expansion of the laser field and lead to a amplification of the field amplitude. Finally, at the peak interaction moment ($t=200\,\text{fs}$, \cref{fig:3}(c)), the laser pulse reaches the cone tip. The synergy of self-focusing and geometric compression results in an extreme field enhancement, with the normalized electric field $E_y$ reaching a peak amplitude nearly 3 times higher than the initial focusing limit. 
\begin{figure}[H]
 \centering
 \includegraphics[keepaspectratio, width=0.48\textwidth]{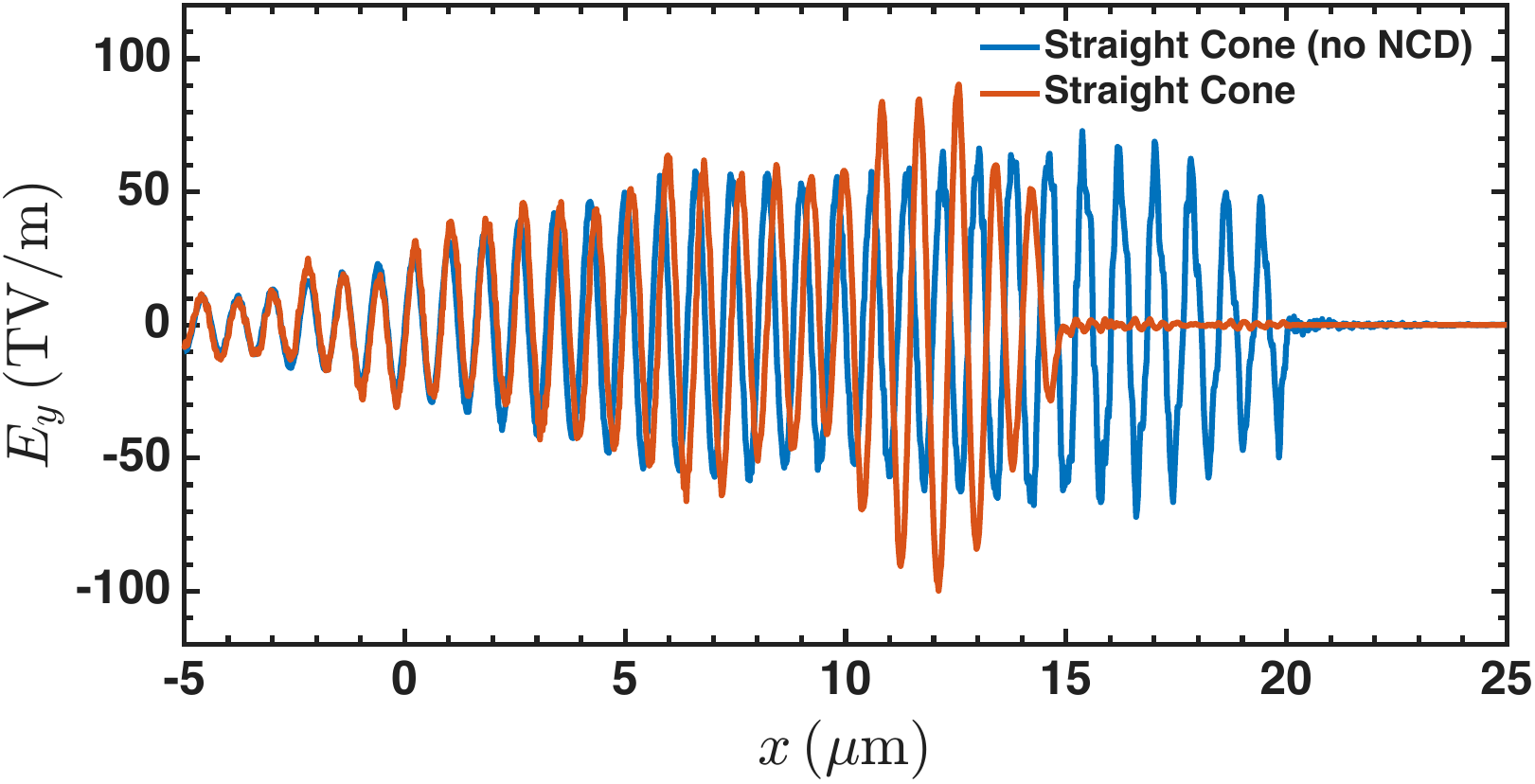}
 \caption{
    Comparison of the on-axis transverse electric field distribution $E_y(x)$ for the Straight Cone target with (orange line) and without (blue line) NCD plasma filling at $t=150\,\text{fs}$. }
 \label{fig:4}
\end{figure} 

\begin{figure}[H]
 \centering
 \includegraphics[keepaspectratio, width=0.46\textwidth]{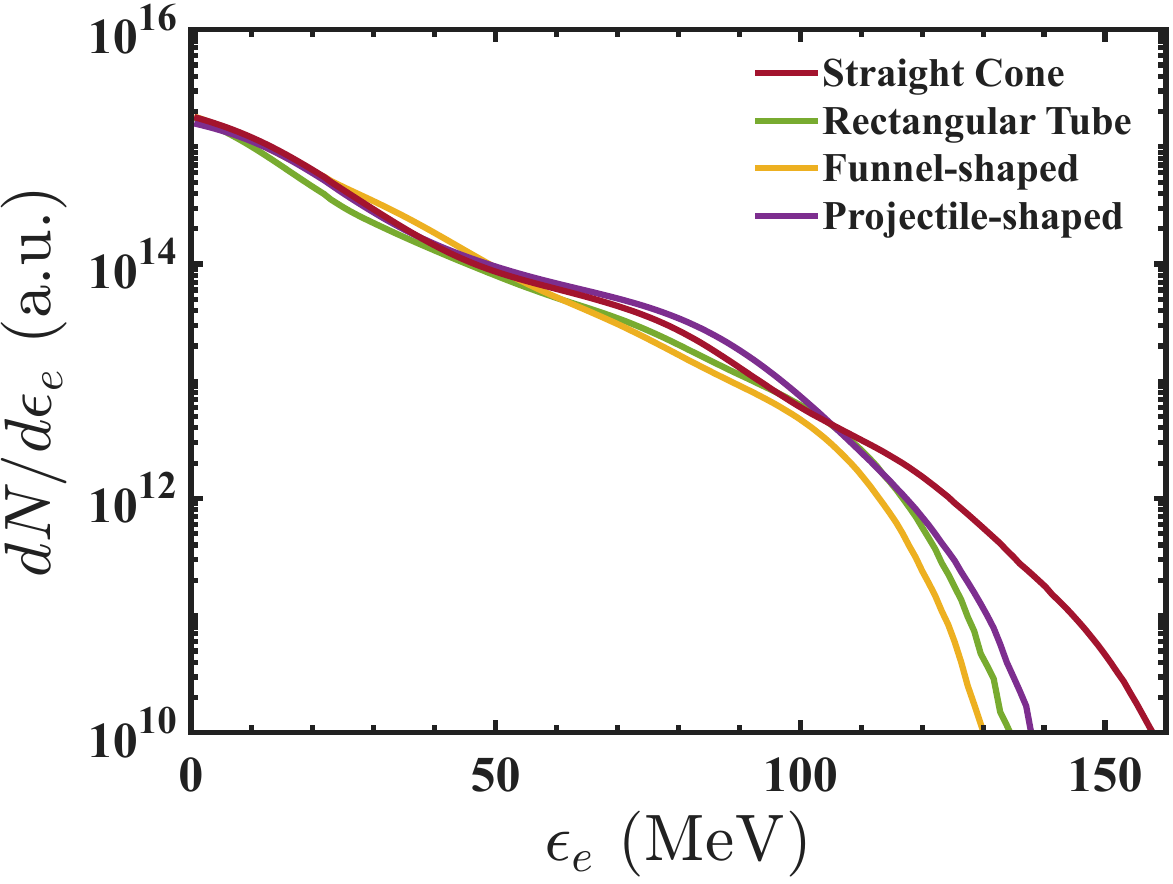}
 \caption{Smoothed electron energy spectra detected behind the target for different geometric configurations at $t=200\,\text{fs}$.}
 \label{fig:5}
\end{figure} 

Furthermore, the propagation of ultra-intense laser pulses through an NCD plasma typically triggers severe parametric instabilities, such as relativistic filamentation\cite{bret2005characterization} and two-stream instabilities \cite{dieckmann2005particle}. In unbounded plasmas, these instabilities can severely degrade the laser envelope and increase the transverse emittance of the accelerated hot electrons. However, our proposed straight-cone geometry effectively mitigates these detrimental effects. As the laser pulse undergoes filamentation or as hot electrons acquire significant transverse momentum ($p_y$), the rigid spatial confinement provided by the solid cone walls forces the transversely scattered laser energy and divergent electrons to be reflected and refocused towards the central axis \cite{snyder2016enhancement, honrubia2017intense}. This mechanical suppression of transverse divergence not only maintains a highly localized and intense accelerating sheath field (as evidenced by the $>100\text{ TV/m}$ peak in \cref{fig:4}), but also effectively converts the deleterious transverse momentum of instability-driven electrons into useful longitudinal refluxing \cite{tresca2011controlling}. Consequently, this structural advantage ensures the high collimation of the accelerated proton beam (\cref{fig:2b}(b)) despite the complex plasma dynamics within the NCD channel.

We compare the on-axis transverse electric field distribution of the NCD-filled straight cone with the vacuum case in \cref{fig:4}. While the electric field in the vacuum target (blue line) remains relatively weak and suffers from diffraction losses near the tip, the NCD-filled target (orange line) exhibits a dramatic surge in intensity, particularly in the range of $x \approx 15\text{--}20\,\mu\text{m}$. The peak amplitude exceeds $100\,\text{TV/m}$, significantly outperforming the vacuum counterpart. This observation aligns with the findings of Horný and Doria \cite{horny2025multi}, confirming that the NCD medium effectively facilitates relativistic self-focusing and pulse shaping, thereby optimizing the laser-to-particle energy coupling efficiency.
\begin{figure*}[t]
 \centering
 \includegraphics[keepaspectratio, width=1.06\textwidth]{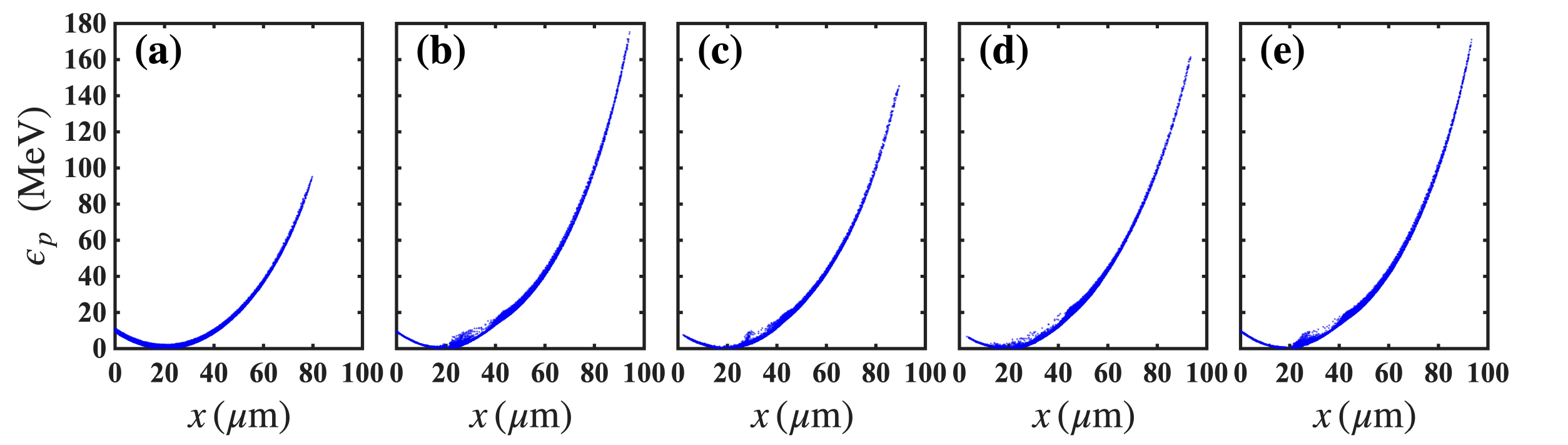}
 \caption{Comparisons of the proton longitudinal phase space distributions at $t=680\,\text{fs}$ for different target geometries: (a) Flat Foil (no NCD), (b) Straight Cone, (c) Rectangular Tube, (d) Funnel-shaped, and (e) Projectile-shaped. The blue dots represent the macro-particles. }
 \label{fig:6}
\end{figure*} 

\Cref{fig:5} presents the smoothed electron energy spectra for the different target geometries. As evident from the comparison, the Straight Cone target (red line) yields a superior production of high-energy electrons compared to the other shapes. While the low-energy populations ($\epsilon_e < 20\,\text{MeV}$) are comparable across the cases, a distinct divergence appears in the high-energy tail. The Straight Cone spectrum extends to a cut-off energy exceeding $140\,\text{MeV}$, whereas the Rectangular Tube spectrum drops off rapidly. The Straight Cone spectrum indicates a higher effective electron temperature ($T_{hot}$). This enhancement is a direct consequence of the synergistic effect of the NCD plasma and the conical geometry.

We also investigate the longitudinal phase space distributions of the protons at $t=680\,\text{fs}$, as shown in \cref{fig:6}. This visualization effectively maps the spatial evolution of the energetic protons and their corresponding energy gain, serving as a direct signature of the accelerating sheath field quality.

Comparing the five target geometries, a distinct variation in the proton acceleration capabilities is observed. In the Flat Foil case (\cref{fig:6}(a)), the protons exhibit a typical thermal-like expansion with a limited spatial extension ($x < 60\,\mu\text{m}$) and a relatively low cutoff energy ($\sim 90\,\text{MeV}$). In sharp contrast, the protons in the Straight Cone target (\cref{fig:6}(b)) cover the maximum propagation distance, reaching up to $x \approx 100\,\mu\text{m}$. Evidently, this geometry facilitates the highest acceleration gradient, enabling the protons to attain a maximum cutoff energy of $\sim 175\,\text{MeV}$—nearly double that of the planar target and  higher than the other shaped targets.

This performance can be attributed to the sustained focusing of hot electrons discussed in \cref{fig:3} and \cref{fig:4}. As noted in similar grooved-target studies, the confinement of electrons leads to a stronger and longer-lived sheath field\cite{khan2023enhanced}. In our Straight Cone case, the geometric effect continuously injects a high-density electron bunch into the acceleration axis, thereby driving the protons to cover the longest distance and gain the maximum kinetic energy among all investigated geometries.

\begin{figure}[H]
 \centering
 \includegraphics[keepaspectratio, width=0.44\textwidth]{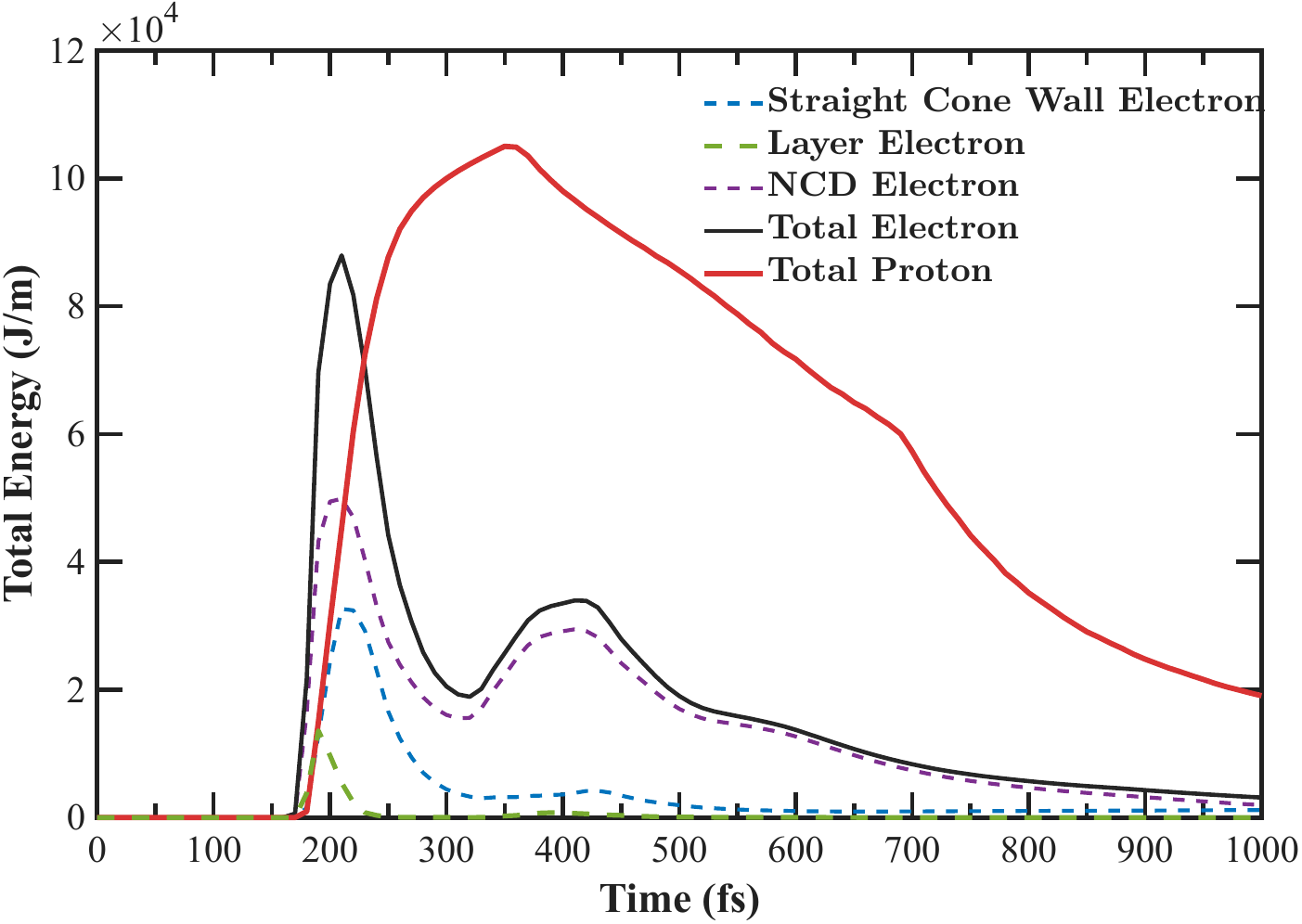}
 \caption{Temporal evolution of the total kinetic energy for different particle populations within the Straight Cone target.}
 \label{fig:7}
\end{figure}

Next, we analyze the time-resolved energy transfer within the target. \Cref{fig:7} presents the evolution of total kinetic energy for protons and various electron populations. The Total Electron energy (black line) rises rapidly during the laser interaction, a trend that is clearly dominated by the NCD electrons (purple dashed line). This confirms that the NCD channel is the primary region for energy absorption via Direct Laser Acceleration. In comparison, the contributions from the Cone Wall (blue dashed line) and the Layer (green dashed line) are significantly lower.

A critical feature observed in the NCD electron profile (purple dashed line) is the emergence of a double-peak structure (or a sustained energy plateau) following the initial laser interaction peak. Instead of decaying monotonically, the energy of the NCD electrons remains elevated and exhibits a secondary maximum at later times ($t > 600\,\text{fs}$). We hypothesize that this phenomenon is a signature of strong electron refluxing. Due to the isolated geometry of the cone, high-energy NCD electrons are confined by the sheath fields at both the front and rear surfaces. Unable to escape immediately, they oscillate (reflux) inside the target, effectively extending their lifetime and contribution to the sheath formation. This refluxing mechanism is highly advantageous for ion acceleration. It maintains a robust and long-lived sheath field at the target rear, preventing the rapid field decay typical of open geometries. Consequently, the Total Proton energy (red line) demonstrates a continuous and steady increase throughout the entire simulation window, verifying that the trapped NCD electrons are efficiently transferring their energy to the protons over an extended acceleration time.

\begin{figure}[H]
 \centering
 \includegraphics[keepaspectratio, width=0.44\textwidth]{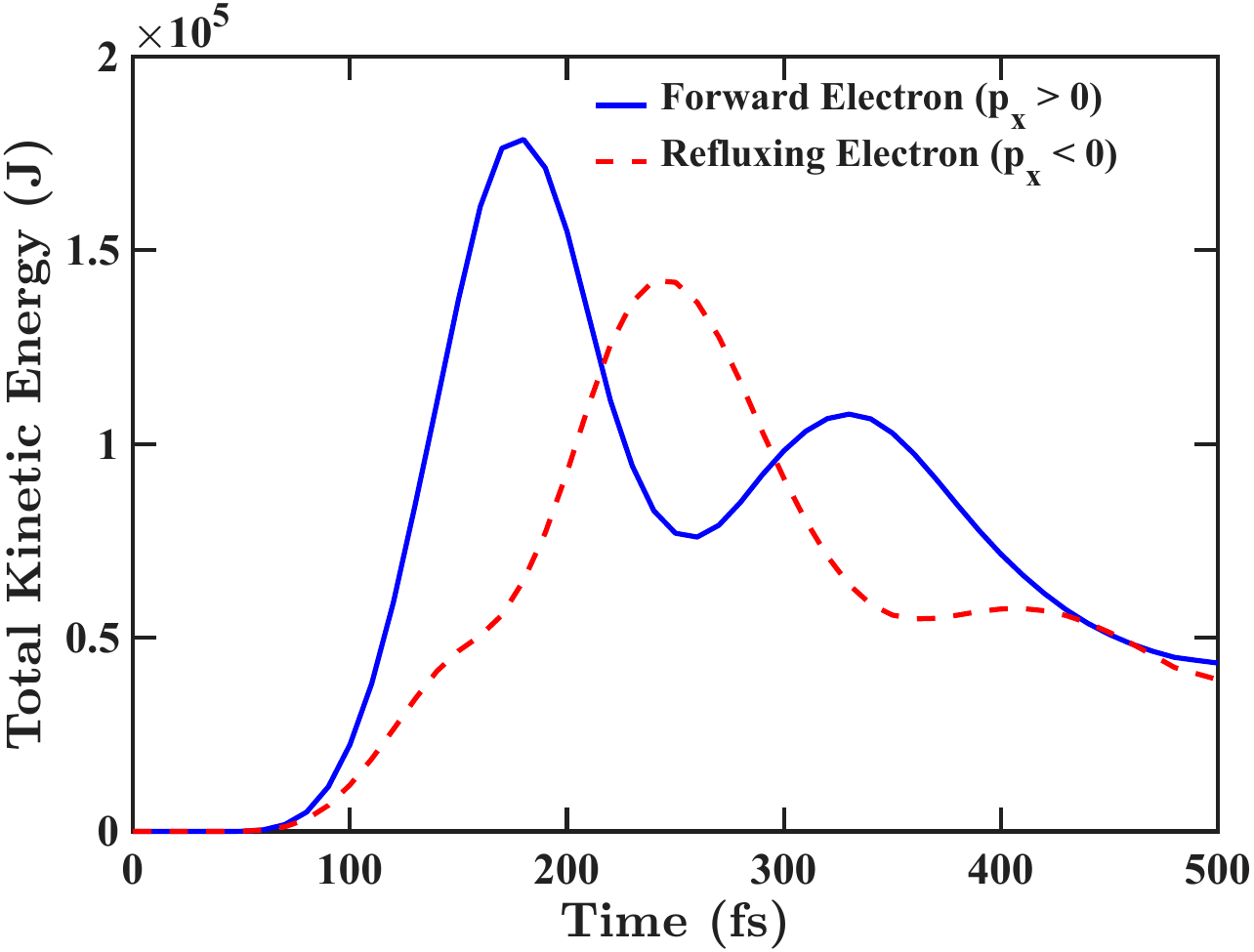}
 \caption{Temporal evolution of the total kinetic energy for forward-moving ($p_x > 0$, blue solid line) and refluxing ($p_x < 0$, red dashed line) electrons.}
 \label{fig:8}
\end{figure}

To explicitly verify the electron refluxing hypothesis, we decomposed the kinetics of the NCD electrons based on their longitudinal momentum. \Cref{fig:8} compares the temporal evolution of the total kinetic energy for forward-moving electrons ($p_x > 0$, blue line) and refluxing electrons ($p_x < 0$, red dashed line). During the main laser interaction phase ($t < 250\,\text{fs}$), the forward electron energy rises sharply, governed by the direct laser acceleration mechanism. Crucially, a significant population of backward-moving electrons emerges with a noticeable time delay of $\sim 50\text{--}100\,\text{fs}$ relative to the forward component.

To quantitatively validate this refluxing mechanism, we can conceptualize the NCD-filled straight cone as a one-dimensional longitudinal electrostatic potential well. The boundaries of this well are defined by the charge-separation field at the target front and the intense target normal sheath field at the rear solid foil. The hot electrons generated via volumetric heating are highly relativistic, propagating with a characteristic velocity $v_e \approx c \approx 0.3\,\mu\text{m/fs}$. Assuming the primary hot electrons are ponderomotively driven forward within the region $x \in [5, 15]\,\mu\text{m}$, where strong relativistic self-focusing occurs, the effective round-trip distance $\Delta x_{\text{eff}}$ required for these electrons to reach the rear sheath, reflect, and re-emerge as a backward-moving population ranges from approximately $15$ to $30\,\mu\text{m}$, depending on their exact birth positions. Analytically, the expected time delay for this refluxing signature is given by:

\begin{equation}
\Delta t_{\text{delay}} \approx \frac{\Delta x_{\text{eff}}}{v_e} \approx 50 \sim 100 \, \text{fs}
\end{equation}
This theoretical estimation is in superb agreement with the transit time and the $\sim 50\text{--}100\,\text{fs}$ temporal delay observed in our PIC simulations.

Post-interaction ($t > 300\,\text{fs}$), while the forward energy stabilizes, the refluxing electron energy does not decay but instead maintains a high plateau, reaching an amplitude comparable to the forward component. Furthermore, the temporal broadening of the refluxing peak (red dashed line) can be attributed to the phase mixing effect of electrons with varying kinetic energies, combined with the dynamic longitudinal expansion of the rear sheath field driven by the escaping protons. This observation provides unambiguous evidence that the energetic electrons are spatially confined and are oscillating between the target boundaries rather than escaping. This mechanism aligns with previous findings in mass-limited targets\cite{tresca2011controlling}, where the transverse refluxing of hot electrons was identified as a critical factor for enhancing the sheath field and boosting the proton cutoff energy. Similarly, in our isolated straight cone geometry, this sustained electron recirculation effectively recycles the kinetic energy, maintaining a robust accelerating field over an extended duration and ultimately driving the superior proton acceleration observed in \cref{fig:5}.

Previous studies by Higginson et al. demonstrated that in the relativistic transparency regime, a jet of super-thermal electrons is generated via direct laser acceleration, which significantly enhances the longitudinal sheath field \cite{higginson2018near}. In our NCD-filled targets, the volumetric heating plays a similar role in generating a population of energetic electrons. However, a critical distinction lies in the geometric confinement provided by the micro-structure walls. Unlike in planar foils where electrons are prone to transverse expansion, the conical walls in our setup mechanically confine the hot electrons. This confinement forces the electrons to oscillate laterally and reflux longitudinally, effectively recycling their kinetic energy to sustain the accelerating sheath field for a longer duration. This mechanism is strongly supported by the double-peak structure observed in the temporal evolution of the electron energy, which serves as a signature of this enhanced refluxing process.

\begin{figure*}[t]
 \centering
 \includegraphics[keepaspectratio, width=1.0\textwidth]{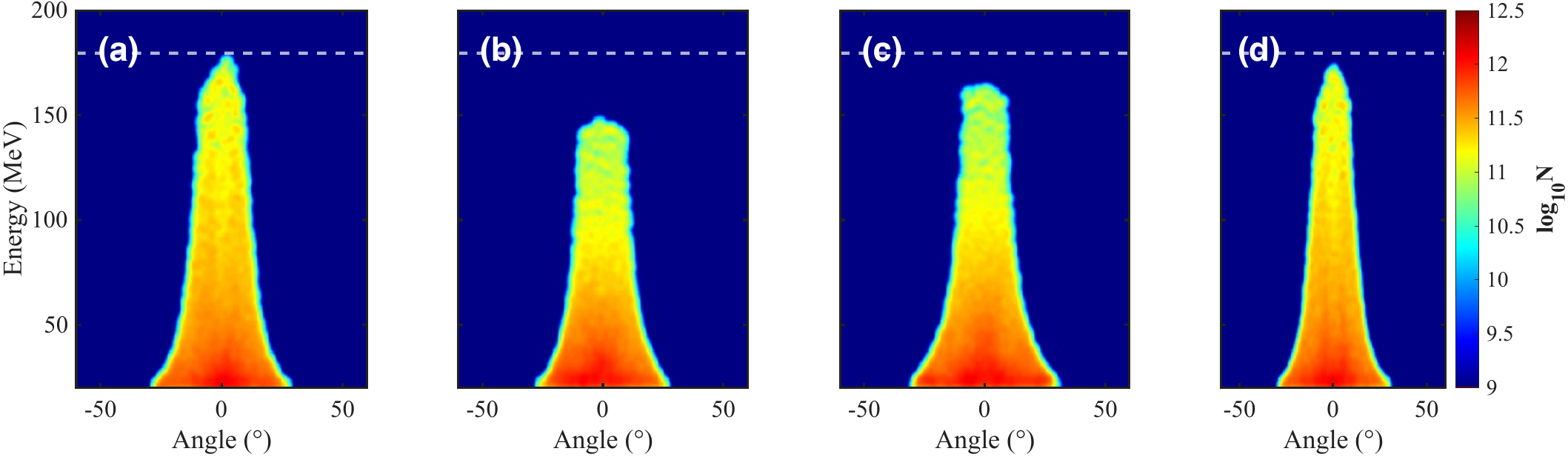}

 \caption{Proton energy-angular distributions at $t = 680$ fs  for different target configurations: (a) Straight Cone, (b) Rectangular Tube, (c) Funnel-shaped, and (d) Projectile-shaped. The color bar represents the macro-particle density on a logarithmic scale.}
 \label{fig:9}
\end{figure*}

To comprehensively evaluate the beam quality and the practical viability of the proposed target designs for applications such as tumor therapy, we further analyze the energy-angular distributions and the overall energy conversion efficiencies. \Cref{fig:9} visualizes the proton phase space for all investigated geometries at $t = 680$ fs. While the projectile-shaped target (\cref{fig:9}(e)) exhibits a relatively sharp energy spread at its tip, the straight cone target (\cref{fig:9}(b)) unmistakably achieves the highest absolute cutoff energy. 

\begin{figure}[H]
 \centering
 \includegraphics[keepaspectratio, width=0.44\textwidth]{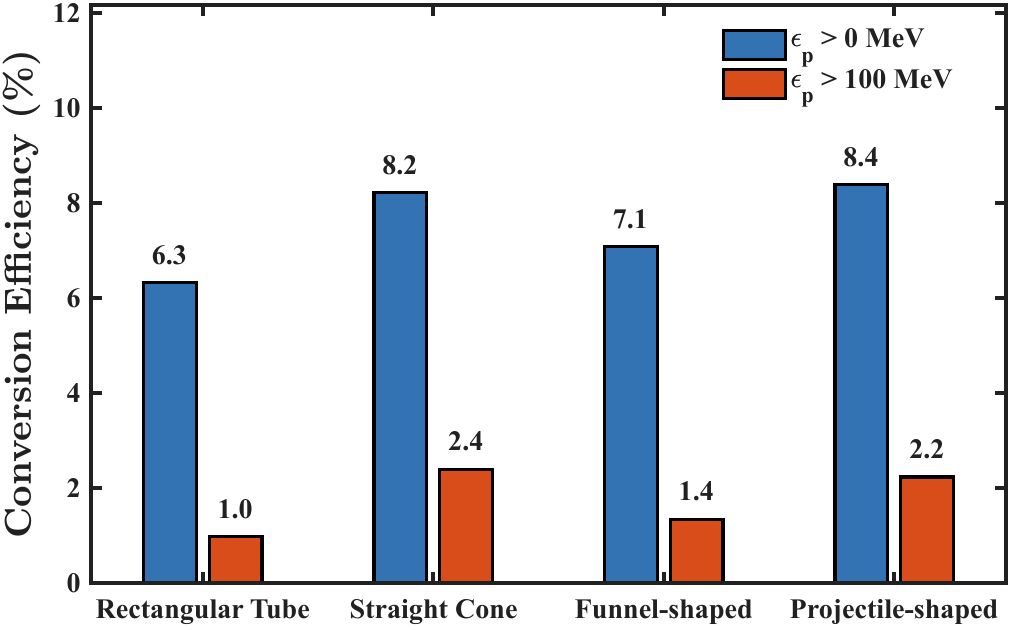}
 \caption{Laser-to-proton energy conversion efficiency for various NCD-filled micro-structured targets. The blue bars represent the total conversion efficiency for all forward-accelerated protons ($\epsilon_p > 0$ MeV), while the orange bars highlight the efficiency specifically for the therapeutically relevant high-energy proton populations ($\epsilon_p > 100$ MeV).}
 \label{fig:10}
\end{figure}

Beyond beam collimation, the laser-to-proton energy conversion efficiency, denoted as $\eta$, is a paramount metric for high-repetition-rate laser facilities. In our 2D PIC simulations, the incident laser energy per unit length is defined analytically based on the Gaussian beam profile.
\begin{equation}
{\text{laser}} = I_0 w \tau \frac{\pi}{4 \ln 2}
\end{equation}
where $I_0$ is the peak laser intensity, $w$ is the focal spot full-width at half-maximum (FWHM), and $\tau$ is the pulse duration FWHM. The total kinetic energy of the accelerated protons is calculated by summing over all forward-moving macro-particles at the target rear.

\begin{equation}
E_{\text{proton}} = \sum_{i} (\gamma_i - 1) m_p c^2 \cdot W_i
\end{equation}
where $\gamma_i$ is the Lorentz factor, $m_p$ is the proton rest mass, $c$ is the speed of light, and $W_i$ represents the statistical weight of the $i$-th macro-particle. The conversion efficiency is thus given by $\eta = (E_{\text{proton}} / E_{\text{laser}}) \times 100\%$. \Cref{fig:10} presents a quantitative comparison of this efficiency across the structured targets. Interestingly, while the projectile-shaped target yields the highest total conversion efficiency for all protons ($\epsilon_p > 0$ MeV) at $8.4\%$, the straight cone target follows closely at $8.2\%$. However, when isolating the therapeutically critical high-energy protons ($\epsilon_p > 100$ MeV), the straight cone unambiguously outperforms all other configurations, reaching an efficiency of $2.4\%$ compared to the projectile's $2.2\%$. This indicates that the continuous and uninterrupted tapering walls of the straight cone not only optimize the total energy absorption but also are the most effective at channeling the absorbed energy specifically into the high-energy proton tail.

\begin{figure}[H]
 \centering
 \includegraphics[keepaspectratio, width=0.44\textwidth]{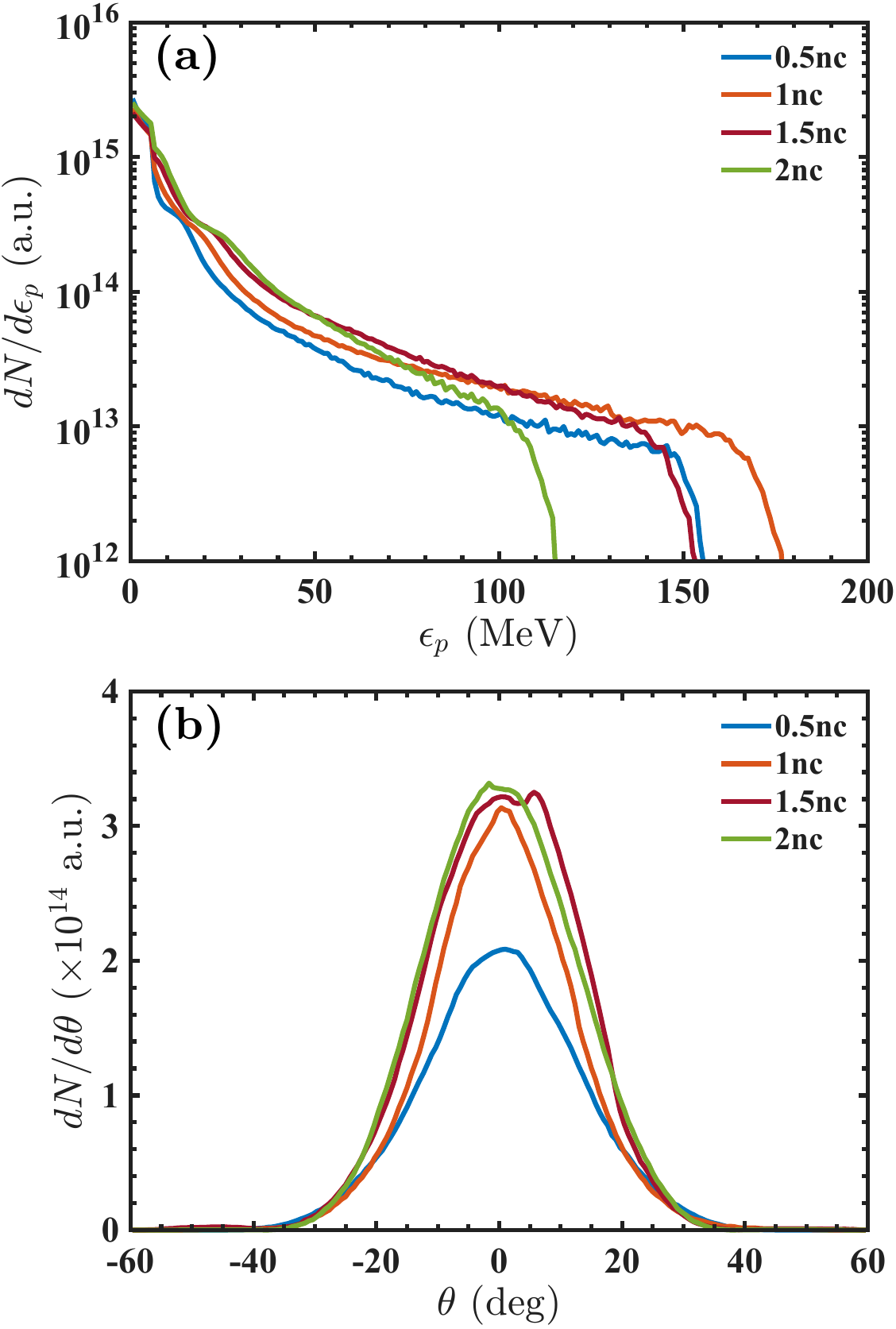}
 \caption{Influence of the initial near-critical density (NCD) on proton acceleration performance at t = 680fs. (a) Proton energy spectra and (b) angular distributions of the accelerated protons for the straight-cone target filled with NCD plasmas at varying initial densities ($0.5n_c$, $1.0n_c$, $1.5n_c$, and $2.0n_c$). The angular distributions in (b) are evaluated for protons with kinetic energies $\epsilon_p > 10$ MeV.}
 \label{fig:11}
\end{figure}

Considering that precisely matching and maintaining an average density of exactly $1.0n_c$ in NCD is technologically challenging in realistic experimental conditions, we performed a parameter scan by varying the initial NCD electron density from $0.5n_c$ to $2.0n_c$. \Cref{fig:11} illustrates the proton energy spectra and the corresponding angular distributions for these varied densities. The results confirm that $n_e = 1.0n_c$ serves as the optimal working point, yielding the highest cutoff energy and the narrowest divergence angle. At this density, the relativistically induced transparency allows the laser to penetrate deeply and undergo self-focusing, while simultaneously generating a massive population of hot electrons via volumetric heating to drive the maximum sheath field. In contrast, when the density drops to $0.5n_c$, the lack of sufficient plasma density limits the total number of accelerated hot electrons, resulting in a weak accelerating gradient. On the other hand, at higher densities ($1.5n_c$ and $2.0n_c$), premature laser energy depletion and enhanced parametric instabilities near the cone entrance prevent the laser from efficiently reaching the tip. This not only degrades the maximum proton energy but also broadens the beam divergence, as clearly evidenced by the flattened angular profile at $2.0n_c$ in \cref{fig:11}(b). This mechanism demonstrates a highly favorable tolerance for experimental inaccuracies. Even with a 50\% deviation from the optimal density (at $0.5n_c$ and $1.5n_c$), the target still successfully produces highly collimated proton beams with cutoff energies exceeding 150 MeV. 
In conclusion, the NCD-filled straight cone emerges as the optimal configuration, outperforming other geometries in both maximum energy reach and beam collimation. This synergistic design provides a robust strategy for generating high-brightness and high-flux therapeutic proton beams.

Our results indicate that the straight-cone target filled with NCD plasma provides the most efficient energy coupling, markedly outperforming conventional flat foils and empty structured targets. This enhancement is primarily driven by the synergy of volumetric heating within the NCD reservoir and the concentrated electron refluxing induced by the cone geometry. 

\section {conclusion}\label{SecIV}
In summary, we have presented a systematic numerical investigation into the enhancement of proton acceleration via the synergistic effect of near-critical density (NCD) plasma and geometric confinement. By leveraging 2D PIC simulations, we decoupled the contributions of volumetric heating and spatial restriction across various micro-structured targets. Our results identify the NCD-filled straight cone as the optimal configuration. This design not only boosts the maximum proton cutoff energy to 181.7 MeV and maintains a highly collimated beam with a divergence of approximately $12^{\circ}$ , but also increases the total yield and flux of high-energy protons. This improvement over conventional flat foils and vacuum micro-structures is driven by the interplay between relativistic laser self-focusing within the NCD channel and the geometric trapping of hot electrons.

A key finding of this study is the counter-intuitive relationship between target complexity and acceleration efficiency. Although we designed complex hybrid targets to improve performance, our simulations show they do not outperform the simple straight cone. This suggests that maximizing geometric continuity may be more advantageous than pursuing complex, multi-stage micro-structures.

Additionally, our analysis reveals a distinct "double-peak" structure in the temporal evolution of the NCD electron energy. This signature provides unambiguous evidence of strong electron refluxing, where energetic electrons are confined by the sheath fields at the target boundaries and oscillate within the cone. This refluxing mechanism effectively recycles kinetic energy, sustaining a robust sheath field ($>100$ TV/m) over an extended duration and preventing the rapid field decay typical of open geometries. Our parametric scans also demonstrate that this synergistic enhancement remains robust across a broad window of NCD densities, while the NCD filling serves as a physical buffer against experimental laser prepulses.

Despite the apparent geometric complexity of the proposed NCD-filled micro-cone, its practical realization is now highly feasible. Ultimately, these findings provide a robust strategy for designing high-brightness, high-energy proton sources tailored for the next generation of Petawatt-class, high-repetition-rate laser facilities. While 3D effects and experimental laser contrast requirements remain factors for future optimization, the synergy of novel target fabrication techniques with our proposed acceleration scheme offers a highly promising pathway toward compact accelerators for medical ion therapy and dense matter radiography.

\section *{Acknowledgments}
This work is supported by the National Natural Science Foundation of China (NSFC) under Grants No.12375240, No.12535015 and the Program of China Scholarship Council (Grant CSC202506040219).

\section *{Conflict of Interest}
The authors declare that there are no conflicts of interest associated with this publication.

\section*{Author Contributions}
\textbf{Cheng-Qi Zhang:} Conceptualization (lead); Methodology (lead); Software (lead); Investigation (lead); Formal analysis (lead); Data curation (lead); Writing -- original draft (lead); Writing -- review \& editing (lead). \\
\textbf{Yang He:} Conceptualization (supporting); Methodology (supporting); Writing -- review \& editing (supporting). \\
\textbf{Mamat Ali Bake:} Formal analysis (supporting); Writing -- review \& editing (supporting). \\
\textbf{Bai-Song Xie:} Supervision (lead); Funding acquisition (lead); 
Writing -- review \& editing (supporting).

\bibliographystyle{unsrt}
\bibliography{ref}
\end{document}